\documentclass[preprint,5p,twocolumn,times]{elsarticle}

\usepackage{graphicx}
\usepackage{amssymb}
\usepackage{amsthm}
\usepackage{upgreek}

\biboptions{numbers}

\usepackage{url}
\usepackage{booktabs}
\usepackage{tabularx}
\usepackage{multirow}   
\usepackage{txfonts}
\usepackage{bbding}     
\usepackage{textcomp} 
\usepackage[hang,small,nooneline]{caption}
\usepackage{fancyhdr}
\usepackage{fixltx2e}
\usepackage[hidelinks]{hyperref}
\usepackage{ulem}   
\newcolumntype{L}[1]{>{\raggedright\arraybackslash}p{#1}} 
\newcolumntype{C}[1]{>{\centering\arraybackslash}p{#1}} 
\newcolumntype{R}[1]{>{\raggedleft\arraybackslash}p{#1}} 

\newcolumntype{M}[1]{>{\centering\arraybackslash}m{#1}} 
\newcolumntype{N}[1]{>{\raggedright\arraybackslash}m{#1}} 

\newcolumntype{Y}{>{\centering\arraybackslash}X} 

\journal{Acta Astronautica}

\begin{document}

\begin{frontmatter}

\title{An innovative concept for the AsteroidFinder/SSB focal plane assembly}

\author[DLR,TUD]{Karsten~Schindler\corref{cor1}}
\ead{karsten.schindler@dlr.de}
\author[DLR]{Matthias Tschentscher}
\ead{matthias.tschentscher@dlr.de}
\author[DLR]{Alexander Koncz}
\ead{alexander.koncz@dlr.de}
\author[DLR]{Michael Solbrig}
\ead{michael.solbrig@dlr.de}
\author[DLR]{Harald Michaelis}
\ead{harald.michaelis@dlr.de}

\cortext[cor1]{Corresponding author}

\address[DLR]{DLR Institute of Planetary Research, Department of Planetary Sensor Systems, Rutherfordstr. 2, 12489 Berlin, Germany}
\address[TUD]{Technische Universit\"at Dresden, Institute of Aerospace Engineering, 01062 Dresden, Germany}

\begin{abstract}
This paper gives a summary on the system concept and design of the focal plane assembly of \mbox{AsteroidFinder/SSB}, a small satellite mission which is currently under development at the German Aerospace Center (DLR). An athermal design concept has been developed in accordance to the requirements of the instrument and spacecraft. Key aspects leading to this approach have been a trade-off study of the mechanical telescope interface, the definition of electrical and thermal interfaces and a material selection which minimizes thermally induced stresses. As a novelty, the structure will be manufactured from a machinable AlN-BN composite ceramic. To enable rapid design iterations and development, an integrated modeling approach has been used to conduct a thermo-mechanical analysis of the proposed concept in order to prove its feasibility. The steady-state temperature distribution for various load cases and the resulting stress and strain within the assembly have both been computed using a finite element simulation.
\end{abstract}


\end{frontmatter}
\thispagestyle{fancy}

\section{Introduction}
\label{sec:introduction}

The primary objective of the \mbox{AsteroidFinder/SSB} mission \cite{Grundmann2009,Michaelis2010} is the detection and follow-up observation of asteroids which orbit the Sun completely interior to Earth's orbit. In contrast to the general progress made in cataloging small bodies in our solar system, no information on the number and properties of these so-called ``Inner-Earth Objects" (IEOs) is currently available, although dynamic models predict a significant population~\cite{Bottke2002}. The major reasons are the obvious limits of ground-based observations resulting from the orbital parameters of IEOs and atmospheric stray light. AsteroidFinder/SSB will be able to access the sky in the VIS-NIR (400\,-\,950\,nm) at solar elongations between 30\textdegree~and~60\textdegree~down to a limiting magnitude of 18.5\,M (\mbox{V-Band}). The satellite will mark the first of a series that is based on a ``standard satellite bus" (SSB) providing a service segment of basic spacecraft functions for a custom payload. It has an envelope of 800\,mm\,\texttimes\,800\,mm\,\texttimes\,1000\,mm, a total weight of 180\,kg and shall be launched to a Sun-synchronous low Earth orbit (LEO, 650\,-\,850\,km) in 2014, probably as a secondary payload. 

\begin{figure*}[tbp]
\centering
\includegraphics[width=0.7\textwidth]{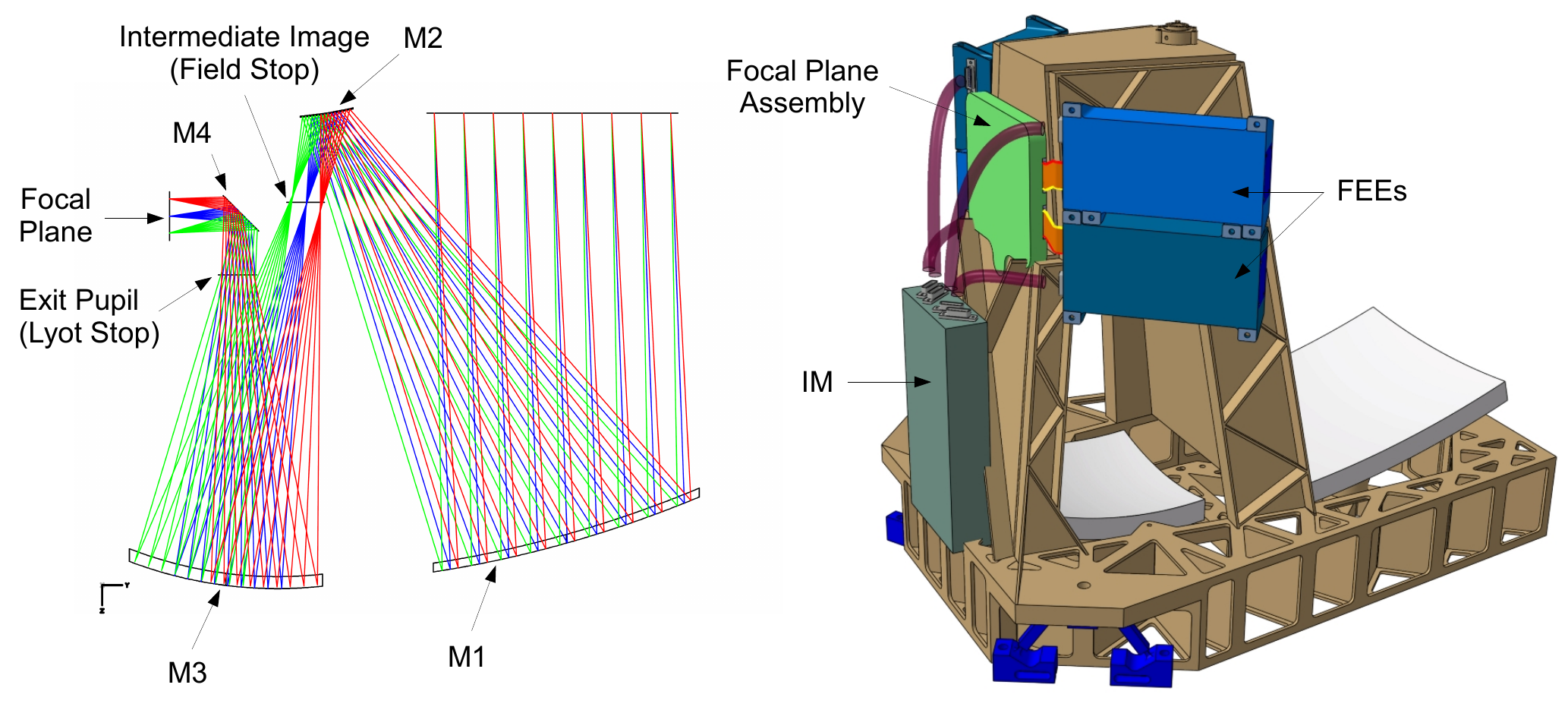}
\caption{Ray trace section and physical layout of the proposed Cook-type TMA telescope on AsteroidFinder/SSB, made entirely from \mbox{HB-Cesic}. The focal plane assembly (FPA) to be discussed in this paper is colored lime green in the right picture. Images by courtesy of \mbox{Dr. Michael Hartl}, Kayser-Threde GmbH, Munich and Matthias Lieder, DLR Berlin.}
\label{fig:AFI_Instrument}
\end{figure*}

The DLR Institute of Planetary Research is responsible for the design, procurement and assembly of the instrument including the attached focal plane assembly (FPA). The proposed telescope design~\cite{Hartl2010b} is a fast (\textless\,f/3.4) off-axis three mirror anastigmat (Cook-type TMA~\cite{Cook1981}) with an additional corrector plate as illustrated in Fig.~\ref{fig:AFI_Instrument}. Both the structure and the mirrors will be made of HB-Cesic, a newly developed hybrid carbon-fiber reinforced silicon carbide (C/SiC) composite ceramic~\cite{Kroedel2007a}. The compact design enables a large field of view (about 2.7\textdegree\,\texttimes\,2.1\textdegree) and provides effective rejection of stray light originating from the Sun, sunlit Earth and other bright sources. The total payload mass of AsteroidFinder/SSB is limited to 28\,kg.

As the satellite bus in its current configuration does not provide a sufficient pointing stability to enable an integration time of 60\,s as necessary to resolve faint objects, electronic image stabilization needs to be implemented on payload level. The FPA will hold four 1k\,\texttimes\,1k \mbox{e2v CCD201-20} backside-illuminated frame-transfer electron multiplying CCDs (\mbox{EMCCDs}) which will operate at 5\,fps to freeze image motion caused by the attitude control jitter of the spacecraft. This also requires a dedicated digital processing unit (DPU) which is able to preprocess, align and accumulate a total of 300 individual images practically in real-time using autonomously selected guide stars to create an equivalent exposure of one minute. The instrument DPU must handle a resulting data rate on the order of \mbox{78\,-\,85\,MBit/s} for each detector.
The FPA electronics architecture is fully redundant. Each \mbox{EMCCD} will have an independent front-end electronic (FEE) chain providing control and analog signal processing and a dedicated power supply. An interconnection module (IM) will act as the interface between the four FEEs and the DPU. It will clock the FEEs based on a quartz oscillator to achieve synchronous read-out of all four EMCCDs within subpixel accuracy.

At the beginning of \mbox{phase B}, it was necessary to develop a detailed design and to define all interfaces of the mission-critical FPA based upon some baseline parameters which had been set earlier in \mbox{phase A}\,/\,$\Delta$A~\cite{Michaelis2009}. This paper summarizes the found athermal design concept and status of the FPA at the time of writing. Innovations compared to previous FPA concepts are the use of a machinable ceramic and an integrated modeling approach with finite elements to speed up the design process. This allowed an immediate thermo-mechanical analysis and a rapid assessment of the temperature distribution and the resulting stress and strain.

\section {Requirements for the focal plane assembly design}
\label{sec:requirements}

The FPA design is constrained by various optical, structural, thermal and electrical requirements. Further aspects to consider are contamination, handling, integration and test. Key design drivers which have been identified are:

\begin{enumerate}
\item
A required operating temperature of the detectors of \linebreak \mbox{\textless\,-80\,\textdegree C} to achieve a sufficiently low dark current
\item
A non-operational temperature limit of \mbox{-95\,\textdegree C}
\item
A very small misalignment tolerance resulting from the fast Cook TMA optic design (\mbox{$\leq\pm$\,40\,\textmu m} total)
\item
A planarity of the active image zones of the EMCCDs of \mbox{$\leq\pm$\,10\,\textmu m} (i.e. 20\,\textmu m peak-to-valley)
\item
The required high stiffness of secondary and tertiary structures (lowest resonance frequency \textgreater\,200\,Hz)
\item
The capability to resist significant mechanical loads, as a variety of launch systems is still under consideration due to the proposed launch of the satellite as a secondary payload
\item
The geometric properties of \mbox{e2v's} \mbox{CCD201-20} EMCCD
\item
Critical thermal control, as cooling of the assembly is only possible by passive means due to the limited power resources of the small satellite
\item
The high read-out rate (5.6\,MHz\,/\,5\,fps) of the detectors
\item
In-flight dark current calibrations, which must be supported by the design of the instrument
\item
Radiation shielding of the FPA electronics and detectors, which needs to provide a shielding equivalent of an 8\,mm aluminum wall shell
\end{enumerate}

The expected temperature difference between the FPA and FEE is on the order of $\approx$\,120\,K. Hence, the thermal resistance of the FEE harness must be as high as possible to minimize the conductive heat flow from the FEE. In contrast, the high read-out frequency requires a low impedance, and signal characteristics recommend a harness length as short as possible. All these requirements need to be achieved by a trade-off between the cross-sectional area of the harness and its length. The team currently favors four flexible polyimide printed circuit boards (PCBs) with copper tracks terminating in Nano-D connectors at both ends to connect the FPA with the FEEs. Flexible PCBs allow us to design a harness with precisely separated tracks that have a cross-sectional area which is significantly smaller than for discrete wires. This is strongly in favor of thermal decoupling and a mechanically flexible connection. 

Likewise, the expected temperature difference between the instrument flange and FPA is also large (current estimate: \mbox{75\,K\,-\,90\,K}). Therefore, the mechanical interface needs to be not only stiff, but must also have the highest possible thermal resistance. This issue will be discussed in Sec.~\ref{sec:mount}.

The high read-out rate of the EMCCDs requires an amplifier in direct proximity to the CCD output. The necessary circuit board will be exposed to temperatures on the order of the detector temperature. This is a significant difference in the FPA design compared to standard camera systems working at lower read-out frequencies, as e.g. instruments in ground-based astronomy. At much lower read-out frequencies, proximity electronic modules (PEM) in the cold environment of the detector are obsolete. Resulting issues which complicate the design of the AsteroidFinder/SSB FPA are the necessity for low temperature resistant components, an appropriate assembly and packaging technology and a qualification of the proximity electronic substrates at temperatures below the requirements of typical MIL specifications~\cite{MIL-STD-883F} and other standards which are commonly used in spacecraft engineering. Survival heaters at the FPA structure will ensure that the proximity electronic substrates will remain at acceptable temperatures during commissioning of the satellite or times when the instrument electronics are switched off for a longer period.

The selected EMCCD has a large number of bond pads at three of its four sides which form the electrical interface to the aluminum nitride high temperature co-fired ceramic (AlN HTCC) substrate underneath. This means the CCD is only buttable at its front face, which results in a large butting loss of about 9 mm in the horizontal direction of the array and 1.5\,mm in the vertical direction (fill factor of the array $\approx$\,70\,\%). A seamless mosaic is not required to fulfill the mission goal, as images from each detector will later be processed individually. However, the array's butting loss raises the requirements applied to the instrument optics in terms of a larger distortion-free field of view.

The design and geometry of the CCD carrier has not yet been finished. The current status suggests that all four EMCCD will share the same (``monolithic") carrier. This is in favor of the requirement that the sensors' surfaces need to be extremely planar ($\leq\pm$\,10\,\textmu m). Still, this already challenging planarity tolerance further reduces the total misalignment budget of the detectors for integration at the instrument to about \textpm\,30\,\textmu m. For each detector, two rows of bond pads located at the adjacent edges of the CCD subcarrier have been defined as the electrical interface to the PEM. This enables a clear separation of responsibilities between the design and production of the CCD package (e2v) and the design and procurement of the PEMs as well as the assembly of the FPA (DLR).

The cold straps which will be attached to the FPA structure and CCD package for passive cooling will be made from high purity, oxygen-free, highly conductive (OFHC) copper braids. The spacecraft radiator will likely be located only about 50\,mm beyond the FPA which results in a very short strap length ($\le$\,80\,mm).

\section {Telescope interface}
\label{sec:mount}
A detailed discussion of the telescope interface was necessary at the very beginning of the design process due to its large influence on the resulting design. It must enable a precise alignment of the detectors during integration and guarantee their position within the given limits throughout the operational life time and temperature range of the instrument. Two solutions well known in opto-mechanical engineering -- a kinematic coupling~\cite{Hale2001,Hale1999} and a flexure mount~\cite{Vokubratovich1988} -- have been studied in detail as a trade-off in terms of their stiffness, minimum size and resulting thermal conductance.

In particular, the telescope interface must have the ability to compensate the difference in thermal expansion between the instrument flange and the FPA without introducing critical thermal stress and strain. The interface should also allow some tilt of the detector surfaces to be able to align them coplanar to the focal plane of the telescope to compensate alignment errors of the package and manufacturing inaccuracies. 

The system's natural frequencies can be obtained by solving an eigenvalue problem. For a first analysis, the system was simplified to a discrete number of degrees of freedom. It was assumed that the telescope and FPA are rigid bodies, which is a fair assumption as both the instrument and the FPA will be made of a stiff ceramic material as discussed later in Sec.~\ref{sec:materials}. 

\subsection {Maxwell mount}
A kinematic coupling is a deterministically constrained system, as six contact points uniquely define the location of the mounted component with respect to its counterpart. Probably its most popular type is the so-called ``Maxwell mount", also known as ``Three-V(ee) Mount". This mount consists of three spheres resting in V- or gothic-arch shaped grooves and can achieve positioning repeatability on the sub-micron scale both in static and dynamic load cases as widely discussed in the literature~\cite{Slocum1988a,Slocum1988b,Zelenika2002}. In fact, repeatability outperforms manufacturing accuracy by orders and is mostly determined by the quality of surface finish of the two components in contact. With a properly applied preload, differences in thermal expansion can be accommodated with minimal stress as the sphere can slide in the groove. Great attention must be paid to the stress concentration at the point contact between the sphere and groove as it can initiate local yielding. However, given the correct choice of materials, wear-in and yielding can be avoided. To predict the stress and deformation at the point contact between the two elastic solid bodies, the non-linear Hertz theory is used. The exact analytical solution as described in~\cite{Boresi2003} has been implemented in a computer algebra system to avoid errors caused by different approximations as discussed in~\cite{Zelenika2002}.

\begin{figure*}[tbp]
\centering
\includegraphics[width=0.6\textwidth]{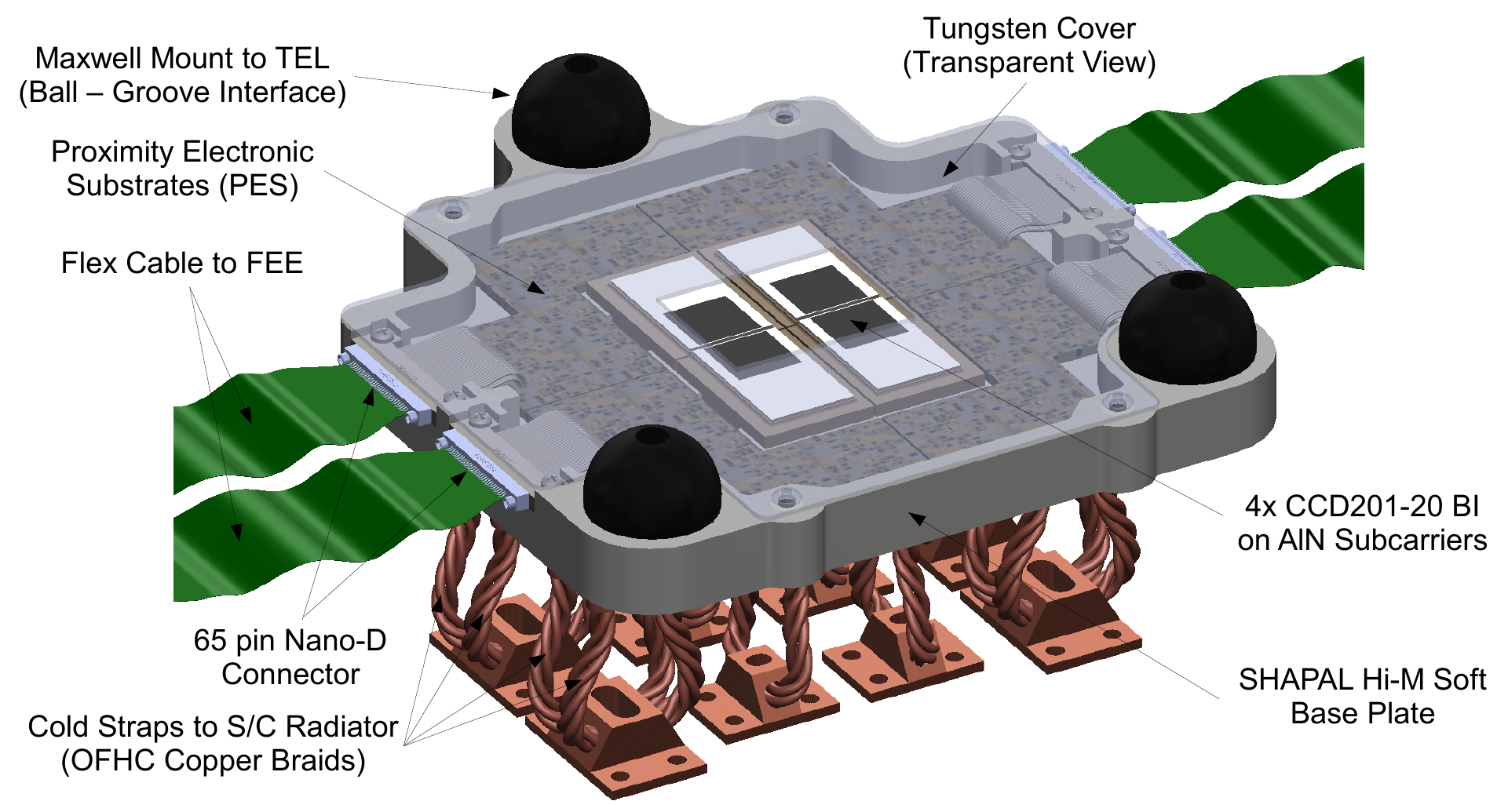}
\caption{The CAD model of the FPA which has been adopted for an integrated thermo-mechanical analysis using finite elements. The tungsten cover has been set transparent for clarity.}
\label{fig:FPA_Concept}
\end{figure*}

\begin{figure}[tbp]
\centering
\includegraphics[width=0.5\linewidth]{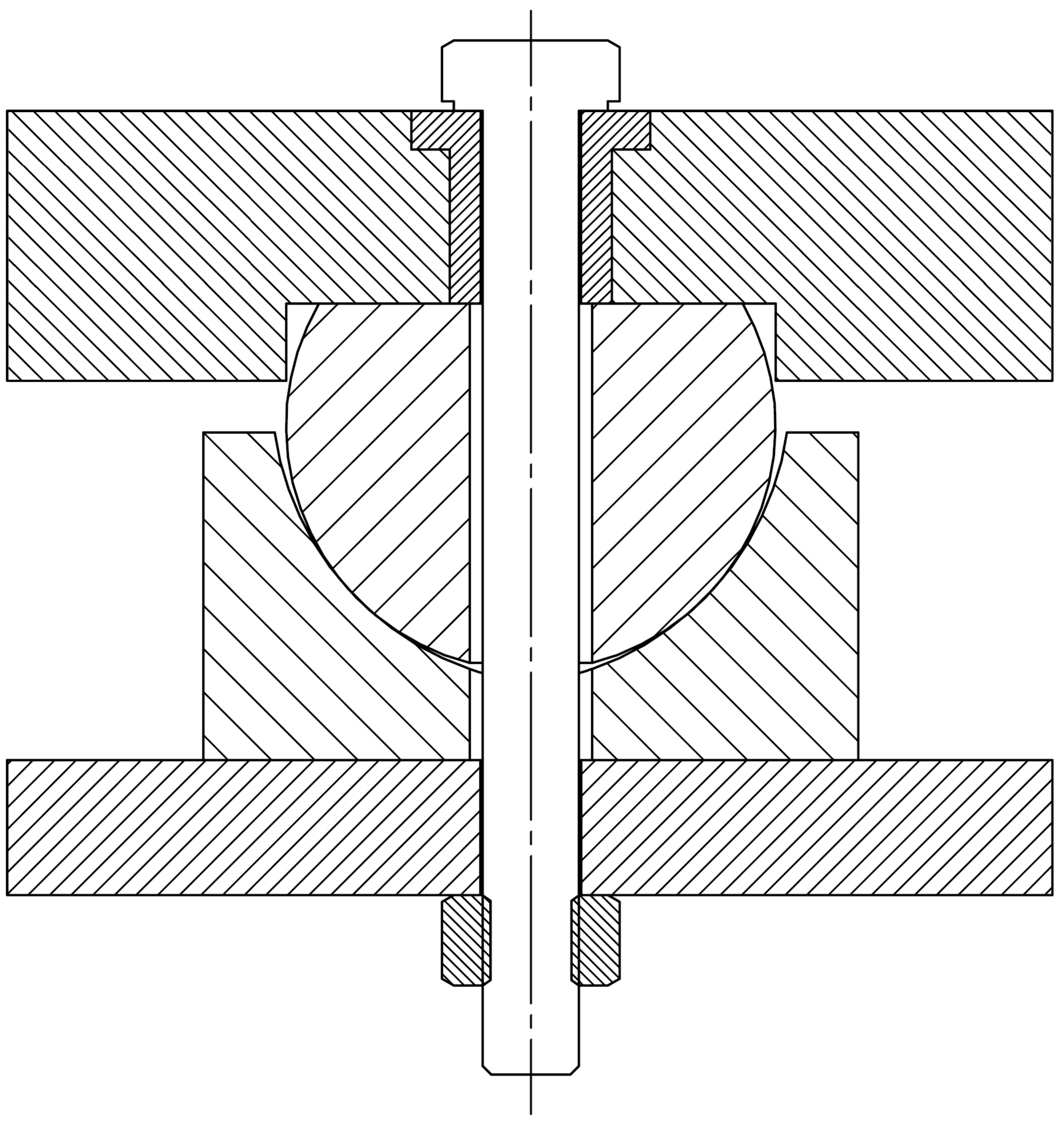}
\caption{Sectional view of the FPA's kinematic mount at the telescope, illustrating one sphere on the FPA resting inside a gothic-arch shaped groove on the telescope flange.}
\label{fig:Kinematic_Coupling}
\end{figure}

An equilateral positioning of the three spheres was not possible as this would significantly increase the size of the FPA. The design resulted in an equal-sided arrangement at angles of 131\textdegree-98\textdegree-131\textdegree~as illustrated in Fig.~\ref{fig:FPA_Concept}. The optical axis must be coincident with the triangle's centroid with the grooves' axes being oriented towards it. This enables uniform distribution of thermally induced stress as well as uniform shrinkage around the optical axis ensuring the detectors remain centered on the optical axis without sliding in x and y. Figure~\ref{fig:Kinematic_Coupling} illustrates the concept of the telescope interface. The stiffness of the coupling was estimated using the approach as presented by Hale~\cite{Hale1999}. The kinematic coupling can be represented by six springs which are tilted by 45\textdegree~to the x-y plane of the FPA and tangentially aligned on a circle centered at the optical axis. 

\begin{figure}[tbp]
\centering
\includegraphics[width=\linewidth]{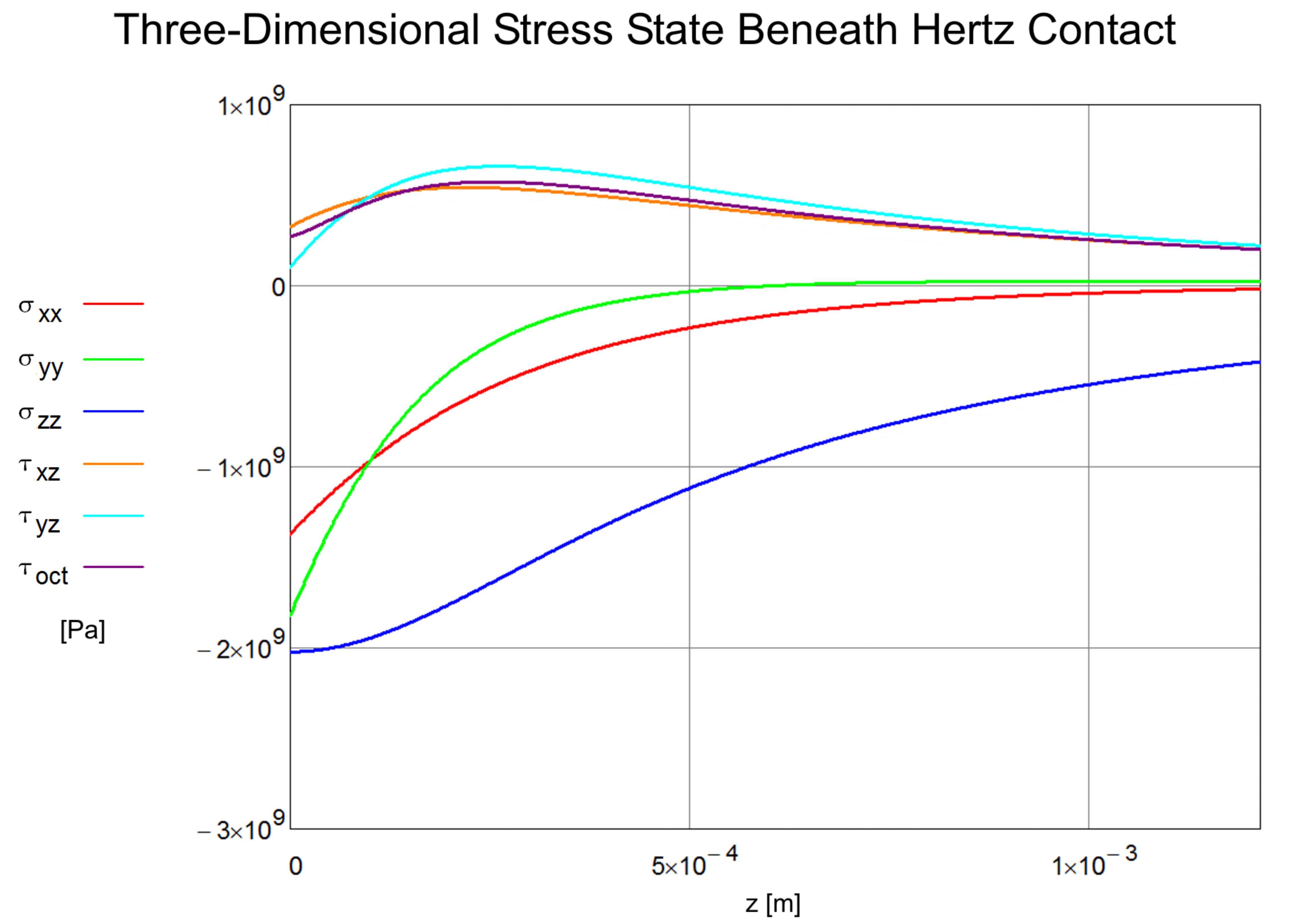}
\caption{Three-dimensional stress state as a function of depth~\textit{z} beneath the center of the Hertz contact between sphere and groove.}
\label{fig:Stress_State}
\end{figure}

The applied preload is one of the main parameters affecting repeatability and stiffness. It is also the essential value to determine if the Hertz contact is still elastically deforming. The preload had to be estimated such that the FPA is securely fastened at the instrument during qualification vibration loads. It can be shown by an analysis of the three-dimensional stress state as a function of depth beneath the center of the contact area (see Fig.~\ref{fig:Stress_State}) that failure is not caused by compressive stress at the surface, but by shear stress which has its maximum at a certain depth inside the solid. 

For metals, yielding happens at this shear stress maximum inside the solid, causing crack growth and pitting. The design and preload of the kinematic coupling must therefore ensure that no stress in any dimension causes plastic deformation, and that especially the shear stress does not initiate yield. This can be checked with a number of yield criteria (e.g. Tresca, von Mises). However, these yield criteria are not applicable to ceramics where it is much more difficult to phrase a sufficient criterion as no yield point exists. Several theories in fracture mechanics try to model the strength of a ceramic. Within this work, empirically found values based on experience obtained from roller bearing development~\cite{Slocum1992_Book} have been used to assess if a ceramic part is safe from fracture, claiming that for a ceramic part in a point contact the limiting factor is its flexural strength \textit{R}\textsubscript{flexure}. Therein, the allowable Hertz contact stress \textit{q}\textsubscript{max} is given as $q_{\mathrm{max}} = {2\,R_{\mathrm{flexure}}}/{(1-2\nu)}$, where $\nu$ denotes Poisson's ratio.

It is obvious that the materials of both bodies in contact need to have a high yield or flexural strength to sustain yielding or failure. For reasons of good repeatability and predictable behavior, both deformation and friction in the contact areas need to be controlled which can be achieved by stiff materials with a high modulus of elasticity and a good surface finish. Moreover, the stiffer the materials of the coupling are, the smaller will be the contact area. This raises the thermal contact resistance between the sphere and the groove, but also the mechanical stresses.

For these reasons, the sphere should be made of a ceramic while the groove should be made of a metal alloy. Silicon nitride (Si\textsubscript{3}N\textsubscript{4}) is lightweight and has a density and a CTE which is well-matched to the selected material of the FPA structure (see Sec.~\ref{sec:materials}). This is important as the sphere will be epoxied to the base plate. Extensive reports of proven sustainability of Si\textsubscript{3}N\textsubscript{4} spheres together with stainless steel can be found in the literature, also because this material combination has been studied extensively e.g. for high performance roller bearings. For this reason, silicon nitride spheres are also relatively easy available. Therefore, Si\textsubscript{3}N\textsubscript{4} was selected as the baseline for the sphere material. The grooves shall be made of a special stainless high alloy steel (1.4534) providing excellent yield and tensile strength while being machinable on typical CNC machines. 

The size of the contact area in the coupling is especially of interest for thermal analysis, as this area mainly determines the contact conductance \textit{k} between the sphere and the groove. Due to the very large contact stress, it is expected that the sphere and the groove are well-coupled (rough order of magnitude estimate: \textit{k}~$\approx$~5000\,W/m\textsuperscript{2}K).

\subsection {Flexure mount}
Flexures could be an alternative to compensate differences in thermal expansion between the FPA and the telescope flange. They are compliant in the radial direction of the mounted element, stiff in all other directions, and have no friction or wear. Flexure mounts are known in many different shapes \cite{Lobontiu2003}. One type widely used in many applications in spacecraft design are bipods as shown for the isostatic mount of the telescope earlier in Fig.~\ref{fig:AFI_Instrument}.
However, in the present scenario, flexure design is driven by two conflicting requirements: high thermal resistance (flexure as long as possible) vs. high stiffness (flexure as short as possible). Moreover, the flexure will undergo a length change which is already large compared to the total misalignment budget of the instrument. In addition, the flexure geometry providing optimal compliance and mass is usually found after several elaborate iterations using FEA. 
The flexure mount has been directly compared to the previously discussed kinematic mount using the same approach as discussed in~\cite{Hale1999}. The mount consisted of three flexure blades, which had been tangentially aligned on a circle centered at the optical axis and aligned in parallel to the optical axis. The flexure mount was then again modeled as a network of springs.

\subsection {Conclusions}

The trade-off study lead to the selection of the ``Maxwell mount", as this mount offers a significantly higher stiffness, higher thermal resistance, reproducible and rather easy alignment and analytically predictable behavior. The arch-shape of the groove relaxes the stress state to an acceptable level in comparison to a V-shaped groove with two flat planes, which is a direct result from Hertz' theory. The sphere-groove point contact is highly isolating from a thermal point of view thanks to the small area in contact, and conductance through the bolt which applies the necessary preload on the mount can be limited by using a low-conductive alloy such as \mbox{Ti-6Al-V4}. Deformation in the Hertz contact needs to be considered in the focus budget, but is predictable: At the necessary preload, the sphere and groove move towards each other by 15.2\,\textmu m, resulting in a deflection of the FPA on the optical axis of the instrument of about 10.7\,\textmu m. As long as the applied torque on the screws is repeatable and fairly accurate, this deflection can be repeated with every assembly. 

\section {Material selection}
\label{sec:materials}

The major goal was to accomplish a passive athermal design of the FPA in accordance to the instrument. Materials have been selected such that the optical performance will be insensitive to the change in temperature from the ground environment to space. Two important constraints which have been set during \mbox{phase A} and have already been mentioned must be considered:
\begin{itemize}
\item
A full ``HB-Cesic"~\cite{Kroedel2007a} (hybrid carbon-fiber reinforced SiC composite) telescope structure
\item
A CCD subcarrier made of a structured aluminum nitride (AlN) high temperature co-fired ceramic (HTCC)
\end{itemize}
AlN has been selected as a package material due to its excellent match in terms of coefficient of thermal expansion (CTE) to silicon~\cite{Chanchani1990} (see Fig.~\ref{fig:CTE_Data}) as it is the ultimate requirement of the FPA design not to stress the detectors. Moreover, it is an electrical insulator which can be structured using different metallization pastes while being highly thermal conductive. To avoid thermally induced stress and strain, the FPA's basic structure (and its components) must now be well matched both to the carrier and the instrument in terms of CTE to achieve great dimensional stability over temperature. Likewise, the structure material needs to have a high thermal conductivity to reduce temperature gradients in the assembly and a high stiffness to save mass. 

An extensive survey on possible candidate materials for the FPA's base plate made soon clear that the best suited candidates having all desired properties are in analogy to the telescope and CCD carrier technical ceramics. Obtained expansion data of AlN and SiC crystals~\cite{Slack1975} and polycrystalline compounds~\cite{Simon1994,Xinetics_2002} as a function of temperature made clear that AlN and SiC ceramics are the materials with the closest match to Si in terms of CTE in a temperature range of about \mbox{150\,-\,300\,K}. At the same time, these materials are highly thermal conductive. In contrast to a sintered AlN ceramic, SiC and C/SiC compounds are electrically conductive and can only be used for electrically passive structures.

Irrespective of these benefits, a structure made of a sintered ceramic is a cost driver due to the required tools and processes. A sintered SiC or AlN part can only be shaped and machined in the state of a ``soft" green body (including pilot holes) making a mold necessary. During sintering, the part will shrink depending on raw material, shape and size, which results in a reduced shape accuracy after firing. Additional material allowance must be considered to compensate this shrinkage. Once sintered, any machining can only be done using special tools (diamond cutting tools, laser) as the fired part is very hard. Due to this reason, tight tolerances in geometry, precise grinding of surfaces and exact final bore hole dimensions are elaborate.

Obviously, it would be desirable to use a ceramic material with comparable properties which can be machined using common metalworking tools (e.g. high speed steel tools, carbide tools) from a raw piece on a common CNC machine. This would also increase cost-efficiency as only a very small quantity of parts will be necessary.

\subsection {Machinable ceramics}
Machinability in the context of ceramics means rather a controlled fracture than chipping as it is the case for machining metal alloys. In fact, three major kinds of machinable ceramics are available on the market and have been investigated: \mbox{MACOR\texttrademark} glass ceramic by Corning Inc. (and third party derivatives such as e.g. MACERITE), boron nitride (BN) compounds with various compositions and \mbox{SHAPAL\texttrademark} \mbox{``Hi-M-Soft"}, an AlN-BN composite by Tokuyama Corporation.

\begin{figure}[tbp]
\centering
\includegraphics[width=\linewidth]{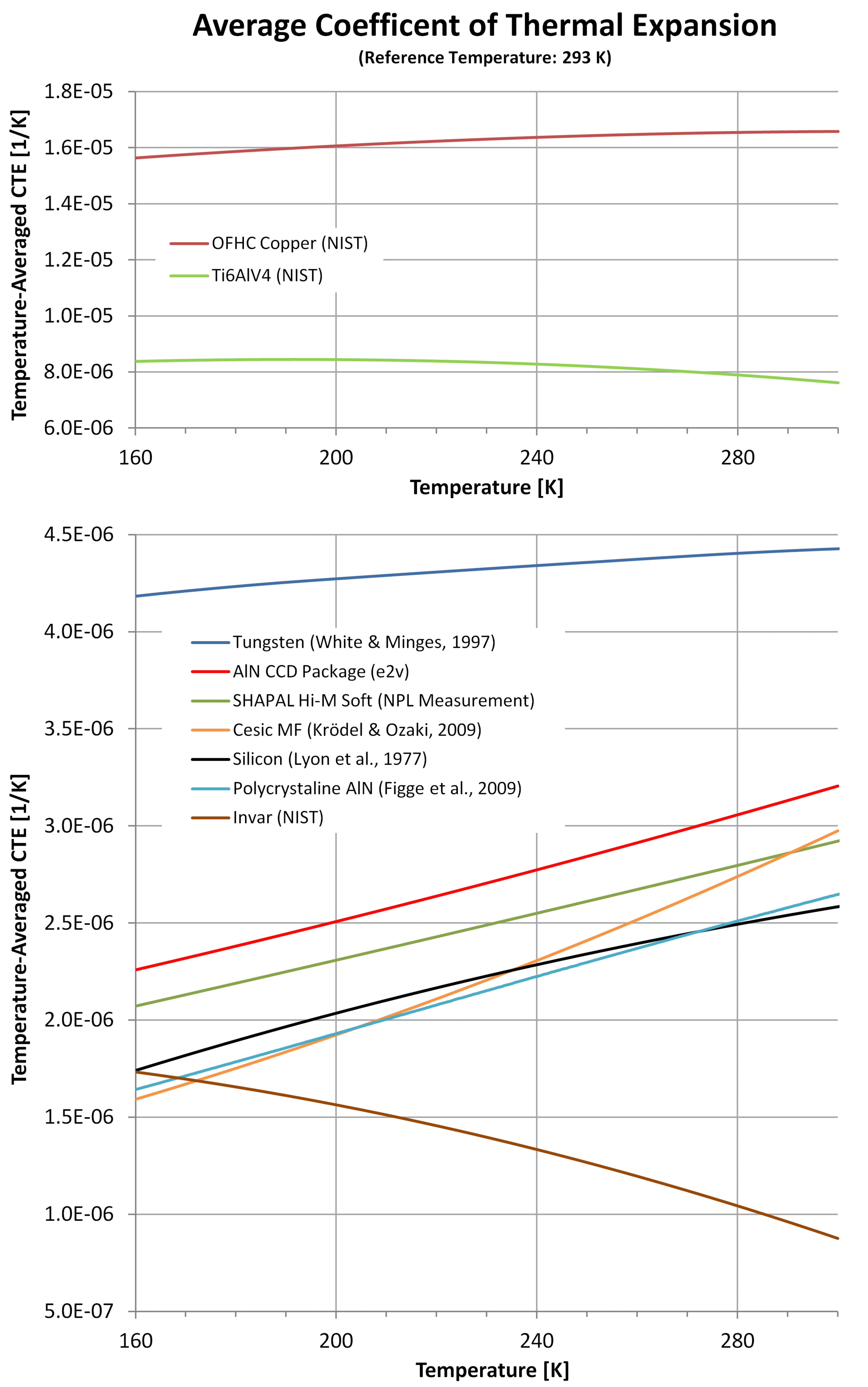}
\caption{Temperature-averaged coefficient of thermal expansion (CTE) of materials which are used in the FPA and instrument design. The used reference temperature has been set to \textit{T}\textsubscript{0}\,=\,293.15\,K. CTE data for the AlN detector package has been kindly provided by Peter Pool at e2v technologies, UK. The data for SHAPAL has been measured by Roger Morrell at NPL, UK on behalf of DLR. For all other materials see~\cite{Lyon1977,Kroedel2009a,Figge2009,NIST,White1997}.}
\label{fig:CTE_Data}
\end{figure}

MACOR was discarded due to its high CTE and very low conductivity. BN ceramics show a significant anisotropy both in mechanical, thermal and electrical properties, as the hexagonal BN crystals align during hot pressing. Depending on the used binder system, BN can also absorb moisture or have a very large porosity which can cause outgassing problems. Therefore, BN ceramics have also been rejected.
SHAPAL \mbox{``Hi-M-Soft"}~\cite{Tokuyama_Product_Info_2011} is a composite consisting of 71\,-\,74\,\%~AlN and 26\,\%~BN~\cite{SHAPAL_Safety_Data_Sheet_2010}. Its zero-porosity and zero-outgassing have been demonstrated by material tests which are documented in ESA's ESMAT database~\cite{ESMAT} making this material applicable for parts in the space environment. Compared to common ceramics, SHAPAL offers an untypically high flexural strength. It also has a high thermal conductivity which is about five times that of Al\textsubscript{2}O\textsubscript{3} of 90\,\% purity, or about 60\,\% of sintered AlN~\cite{Kyocera_MaterialProperties2006}. Due to its high AlN content, its CTE is almost perfectly matched to a sintered AlN ceramic as the CCD subcarrier, while its BN content and proprietary production process enable machinability with common tools. The AlN-like CTE at low temperatures has been verified by an expansion measurement using a dilatometer at the National Physical Laboratory (NPL, UK) on behalf of DLR. The results are illustrated in Fig.~\ref{fig:CTE_Data} in comparison to other materials which find use in the FPA and instrument. SHAPAL's machinability has been demonstrated in a first test. Based on these advantageous properties, vacuum compatibility and low CTE which is optimally matched to AlN and \mbox{HB-Cesic}, it has been selected for the main FPA structure. 

It should be mentioned that SHAPAL has anisotropic properties as it is hot-pressed and likely develops some microstructural texture due to its BN content. To evaluate the extent of this anisotropy, further measurements would be necessary. However, it is believed that this anisotropy is far less severe than for BN solids and probably not of interest for the present application.

\subsection {Metals}

Apart from technical ceramics, only a very limited number of materials have a CTE close to C/SiC, AlN and SHAPAL. 

The commercially most important low expansion alloy family are nickel-iron alloys (\mbox{Fe-Ni}). A nickel content of 36\,\% leads to a minimum CTE. This \mbox{Fe-Ni36} alloy has been named ``Invar" as the size of an Invar sample is practically invariable, however only over a limited temperature range around room temperature. The CTE of Invar at this minimum is about 1/10th of steel. Important derivatives of Invar are ``Free Cut Invar" containing about 0.2\,\% selenium to produce an Invar alloy with a better machinability, ``Super Invar 32-5", a ternary iron-nickel cobalt alloy (\mbox{Fe-Ni32-Co5}) with a CTE of less than half of Invar around room temperature, and Kovar, also a ternary iron alloy (\mbox{Fe-Ni29-Co17}) which is used for sealing applications as its CTE is matched to some popular glass and ceramic materials.

However, Super Invar and Kovar undergo a phase change at temperatures below -50\,\textdegree C~\cite{Yoder2006} or -78.5\,\textdegree C~\cite{Shriver1986}, respectively and are therefore not applicable. In addition, the CTE of Invar also depends much on its heat treatment and applied machining\,\cite{Yoder2006}, so available data need to be carefully reviewed. Invar is also not corrosion resistant, but corrosion-resistance can be achieved by chrome plating. Due to its low thermal conductivity which causes a large temperature gradient, Invar has been discarded for the structure material, but will find use for inserts and connector housings due to the well-matched CTE.

Among pure metals, tungsten has the lowest CTE, followed by molybdenum. Both are very good thermal (and electrical) conductors. Tungsten and its alloys have been discarded for the structure due to their very large density and weight; it is also known to be very difficult to machine. However, a sheet of tungsten will serve as a protective cover on the FPA providing effective radiation shielding. This cover also protects the highly sensitive equipment during handling as it can be easily sealed. 

\mbox{Ti-6Al-V4} has been selected as the material of all fasteners as it has sufficient low-temperature toughness, high tensile strength, low density and one of the lowest conductivities of alloys used for fastener in space applications, which is of great importance for the telescope interface.

\section {Characteristics of the focal plane assembly design}
\label{sec:Specifics}

Designing with ceramics is different as with ductile metals, and some design constraints related to the characteristics of ceramics have to be considered. The main issue is brittleness, as there is no yield and hence stress relief in locally overstressed areas to avoid fracture. The low temperature of the FPA causes additional problems for the design of the proximity electronic modules which need to be addressed. 

\subsection{Engineering specifics of ceramics}

Literature recommends a statistical fatigue analysis to take into account scattering mechanical strength and material properties of ceramics~\cite{Tietz1994,Fischer1992} as caused by already small differences in composition (e.g. chemical purity of raw powders, grain size, homogeneity, additives) and production processes. One suggested approach is to use a survival probability factor which is estimated using the Weibull distribution of the strength values instead of the usual factor of safety known from the design with metal alloys. However, statistical analysis requires detailed knowledge of the ceramic at hand which can usually only be obtained from individual tests of many samples. Often, test data are not available or can not be acquired. This was also the case for AlN and SHAPAL, where the Weibull modulus and fracture toughness are unknown.

Despite of its drawback for ceramics, the common safety factor approach has been used. Safety factors as known from the design of metallic structures (1.1\,-\,2.0) are not applicable to ceramics. The ECSS standard~\cite{ECSS-E-ST-32-10C} recommends a factor of safety (FOS) of 2.5 for ceramic parts undergoing testing and 5.0 for ceramic parts whose design will be verified by analysis only. This large FOS shall cover all uncertainties involved and is easy to apply in a familiar way, although it can sometimes lead to over-engineered parts. 

Besides these rules, some design guidelines as given in~\cite{Tietz1994} have been considered for a first design of the FPA. To avoid weakening of the structure, the minimum wall distance of a hole should be larger than 1/4\textsuperscript{th} of the drill depth or plate thickness for plates up to 10\,mm. The edge distance in a row of bore holes to each other should be larger than twice the hole diameter. Stress-raising sharp edges shall be avoided as well as sudden changes in cross-section due to their crack affinity, hence all edges should have a radius as large as possible. Radii smaller than 0.5\,mm can likely not be manufactured. Punctual force application (e.g. by bolt heads) and notches have to be avoided. The thickness of a ceramic plate which is connected to another part using a bolt should also be at least 0.6\texttimes{} the wrench size of the hexagonal bolt. 

\subsection{Proximity electronic module (PEM) design}

Four Z-shaped proximity electronic modules (PEMs) enclose the CDD packages completely (cf. Fig.~\ref{fig:FPA_Concept}). The preliminary board layout resulted in substrate dimensions having a leg length of about 60\,mm and a width of 15\,-\,20\,mm. Due to this large size, the absolute contraction of the substrate between the cure temperature of adhesives (on the order of 70\,-\,120\,\textdegree C) and the non-operating temperature of the assembly becomes critical not only in terms of reliability of packaging and interconnections, but also in terms of the expansion mismatch to the \mbox{SHAPAL} structure. An elastic mounting is required to compensate the difference in expansion to the basic structure, which is contradictory to the required rigidity as necessary for wedge bonding the PEM to the CCD carrier and to mitigate the occurring vibrational loads during launch.

Standard FR4 and polyimide substrates for the PEMs have been discarded as the mounting of assembled electronic components would likely fail already due to the very large temperature range and fatigue caused by thermal cycling. Composite substrates such as copper-invar-copper (CIC) and copper-molybdenum-copper (CMC) have not yet been considered. The team currently prefers thick-film hybrids as those appear to be a much better solution at very low temperatures~\cite{Kirschmann2001}. The typical hybrid substrate on the market is Al\textsubscript{2}O\textsubscript{3} (alumina) which is available in different purities, while AlN finds mainly use in niche applications which require good heat spreading. Due to the optimal CTE match to SHAPAL, an AlN thick film hybrid has been set as a baseline for this work, which shall be glued to the base plate using a low temperature resistant thermally conductive epoxy. After being glued in place, the assembly is also rigid enough to allow for wedge bonding to the CCD subcarrier. The team currently investigates the technical feasibility of AlN thick film hybrids which need to be manufactured according to space qualification standards.

From the point of project management it appears disadvantageous to directly bond a flexible PCB (flex cable) with a pre-defined fixed length to the PEM ceramic substrate. For greater flexibility, a 65-pin Nano-D connector has been chosen as a universal electrical interface of the PEM to the FEE. This still enables a modification or optimization of the FEE harness at a later stage in the project.

\subsection{Mass and size estimates}
The current mass estimate of the FPA which considers a monolithic CCD carrier is 1317\,g without margin. Adding a system margin of 20\,\% as usual at this project phase results in a total weight of 1580\,g. The FPA's external dimensions can be inscribed into a box sized 181.5\,mm\,\texttimes\,178\,mm\,\texttimes\,40\,mm. The large maximum extent in \mbox{z-direction} is mainly caused by the spheres of the kinematic mount and the cold strap interface sticking out of the basic structure. The major part of the structure is of course much thinner (7\,mm or 14\,mm at the outer rim). Still, a structure optimization needs to be done in a future work as soon as all boundary conditions of the CCD package and PEM are fixed in order to save some percent of the assembly weight.

\section {Integrated thermo-mechanical analysis using finite elements}
\label{sec:FEA}

The major interest of this work has been to investigate the temperature distribution and thermal gradient among the FPA and to obtain a first estimate of the resulting thermal stress and strain in the assembly. The displacement of the CCDs on the optical axis due to the deformation of the FPA at operating temperature needs to be estimated to prove the feasibility of the selected design approach with respect to the small available misalignment tolerance. To allow for a rapid iteration of the design, it has been important to reduce modeling efforts and to have a tool at hand which allows integrated thermal and structural modeling from one source model. As a finite element (FE) model can be used both for structural and thermal modeling, it has been the obvious choice to tackle both tasks at once. A separate finite difference (FD) model for thermal analysis has been discarded although it would have been much smaller and significantly more efficient in terms of computing time. 

A significant drawback is the correlation of the FE model's accuracy with its mesh size. Especially curved surfaces, cut-outs and changes in section require a fine mesh to be accurately represented. Moreover, it is not possible to represent parts of minor interest with just a single node as in FD. Hence, a FE model usually consists of a significantly larger number of elements to converge to a reliable result than a FD model~\cite{Handbuch_Raumfahrttechnik_2008}. Moreover, the temperature is calculated at all nodes at the element's vertices, which leads to a density of temperature reference points which is often not required to get an idea of the overall temperature distribution within a part.

\subsection{Modeling approach}

A finite element model has been implemented in \mbox{SolidWorks} \mbox{Simulation 2009 SP3.1} as this provided some key benefits. Due to the seamless integration of the simulation add-in in \mbox{SolidWorks 2009 SP5.1}, a simplified model is easily derived and maintained as a sub-configuration of the detailed CAD \mbox{3D-model}. The CAD model was mandatory to derive a set of part and assembly drawings as well as information on the assembly's mass and moments of inertia, which are two important inputs for the telescope interface trade-off study discussed earlier in Sec.~\ref{sec:mount}. Moreover, thermal and structural modeling can be done on the same CAD model using the same software, which is especially advantageous in a design phase as early as in this work when properties of the model are still uncertain and subject to changes. 

As CAD, thermal and structural modeling can be done in the same software, it is not necessary to continuously update a representative FE and FD model in third party software for thermal and structural modeling. Within SolidWorks Simulation, the estimated temperature distribution can be used directly as an input for a static analysis which assumes deformation is still elastic. This also avoids all problems related to data exchange between different software packages, different levels of simplifications in geometry, different mesh sizes, interpolation accuracy (e.g. when a coarse mesh of estimated temperatures is projected on a fine mesh for structural analysis) and different data format standards. 

The detailed volume model as presented earlier in Fig.~\ref{fig:FPA_Concept} has been simplified in terms of geometry to reduce mesh size and computational efforts. All fillets and radii have been removed from each part, except for a few very large radii at the FPA structure in proximity to the spheres. The radiator and telescope plane have been included to be able to model conductive and radiative heat flows in interaction with these subunits or towards space (\textit{T}\,=\,3\,K). The resulting FE model for thermal analysis consisted of 116375 elements with 218107 nodes. Different mesh sizes have been applied to different parts and features as necessary.

The model includes contact conductance at component contacts, while radiative heat transfer can be easily modeled even on advanced geometries. However, this comes at the expense of a great increase in computing time as the preprocessor needs to calculate view factors between corresponding finite elements using methods known from ray-tracing (Monte Carlo method) to be able to model surface-to-surface radiation. The FE model contains surface-to-surface radiation between the telescope dummy plate, grooves and all relevant surfaces on the FPA (tungsten cover, uncovered base plate surface, active CCD areas, uncovered parts of CCD carriers and non-active CCD areas, spheres). In addition, the software supports the definition of temperature-dependent material properties as provided in Fig.~\ref{fig:CTE_Data}.

All simulations have been done on an Intel Xeon X5670 workstation (6\,\texttimes\,2.93\,GHz 64\,bit CPU cores, 12\,GB RAM) which could cope with the model presented in this section in a reasonable time frame.

\subsection {Assumptions and boundary conditions}

Thermal contact resistance results from the fact that only microscale surface peaks of materials are in contact. Conductive heat flow at surface contacts is a complex problem depending on many factors such as surface roughness and waviness, use of a filler medium as well as Young’s modulus and surface rigidity of the materials in contact, both determining the extent of elastic and plastic deformation of roughness peaks~\cite{STCH_2002}. No common theory for the calculation of contact resistance exists. Usually, values for contact conductance between two surfaces are based on established estimates from engineering experience which has been obtained in previous projects (as they do in the analysis within this work) or tests on a case-by-case basis, as it is practically impossible to determine contact conductance values analytically. Also, no measurements or recommended values for ceramic-metal or ceramic-ceramic contacts exist in the literature.

Statistically, the number of contacts does increase with decreasing surface roughness. For metals, an increasing surface pressure results in an increased number of yielding peaks leading to a larger contact area. However, as a ceramic cannot yield, an increasing surface pressure has little effect especially for a contact between two ceramic parts. Hence, surface finish will play a major role for heat transfer, and soft filler materials which can deform and yield and therefore wear into the surface roughness of the ceramic part will become important for the cold strap interface. When an adhesive is used for bonding two parts, experience teaches that it does usually wet only a fraction ($\upeta \approx 25-40\,\%$) of the surfaces in contact, significantly decreasing the effective contact area for conductive heat flow.

A short parameter sensitivity study revealed which thermal contact resistances are critical for the analysis. An equivalent network of thermal resistors has been derived to gain a better understanding of conductive heat flows in the assembly and to assess critical parameters in the thermal model. The steady-state temperature distribution has been analyzed for five different representative scenarios: three operational (hot, typical, cold) and two non-operational (hot, cold) cases. The three operational cases differ only by the estimated heat dissipation of the electronic components (CCD, PEM) which are directly integrated on the FPA. The two non-operational cases represent the scenario that all instrument related subunits are switched off for a longer period of time while the SSB also dissipates only a minimum of heat. The non-operational hot case considers the implementation of an active heater while the non-operational cold case represents the worst-case scenario without any thermal control available. The operational hot case has been used to define the size of the radiator and to estimate the maximum temperatures of the FPA components. 

A meaningful evaluation of thermal stress and strain depends on reliable material property data over temperature. CTE data between 150 and 293\,K could be obtained for all materials in the assembly (cf.~Fig.~\ref{fig:CTE_Data}). Information on thermal conductivity over temperature is missing for all ceramics and had to be modeled using a constant value as specified by the individual supplier, however the resulting modeling error is likely well within the inaccuracy of the thermal analysis.

Due to the large temperature difference between the telescope and FPA ($\ge$\,70\,K), the radiative heat flow from the telescope structure must be considered for thermal control of the FPA in addition to the conductive heat flow which enters the FPA through the Maxwell mount. Obtaining reference values for the emissivity $\varepsilon$ of certain materials from literature -- especially for ceramics -- is again difficult. In general, emissivity depends much on surface condition and surface finish. Ceramics tend towards large emissivity values (on the order of $\varepsilon$\,=\,0.8) while pure metals such as tungsten and metal alloys tend towards low emissivity values (on the order of $\varepsilon$\,=\,0.05\,-\,0.1).

\subsection{Results}

The derived equivalent electrical network of thermal resistors has been modeled in \mbox{SPICE} to crosscheck the results which have been obtained from the FE simulation. Both models produced consistent values, which proved the reliability of the FE model used for steady-state thermal analysis.

\begin{figure*}[tbp]
\centering
\includegraphics[width=0.7\textwidth]{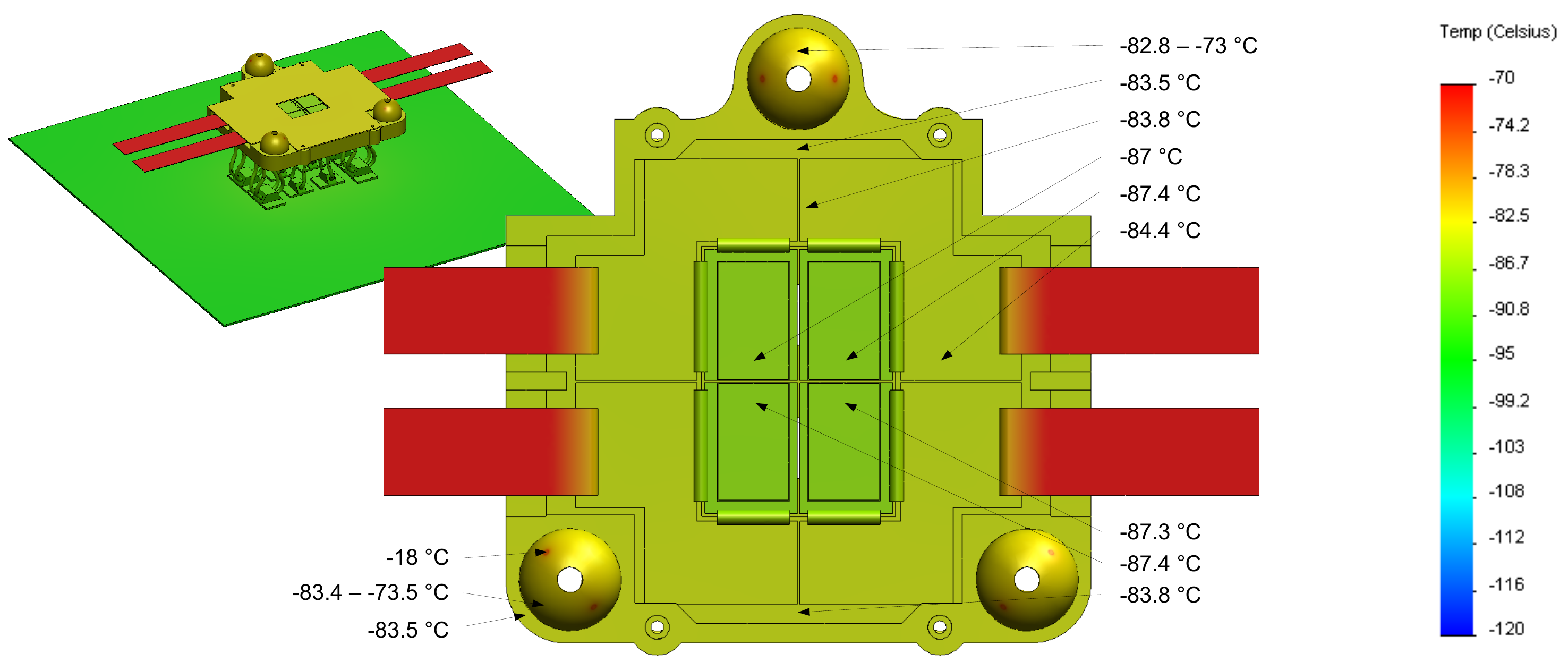}
\caption{Resulting temperature distribution on the FPA in a typical operational load case. The tungsten cover has been masked out in the front view for clarity.}
\label{fig:Temperature_Map}
\end{figure*}

The results of the analysis make clear that the majority of the applied heat flows emerge from external subunits. This is typical in low-temperature and cryogenic engineering. In a nominal operational scenario, just about one-third of the total heat flow is dissipated by electronic components on the FPA, while two thirds still emerge from conductive and radiative heat flows from the telescope and front-end electronics. This emphasizes the need of a careful accommodation of all subunits and a well-studied structural and thermal design of the whole instrument. Figure~\ref{fig:Temperature_Map} gives an example of the estimated steady-state temperature distribution on the FPA for the nominal operational load case.

\begin{figure}[tbp]
\centering
\includegraphics[width=\linewidth]{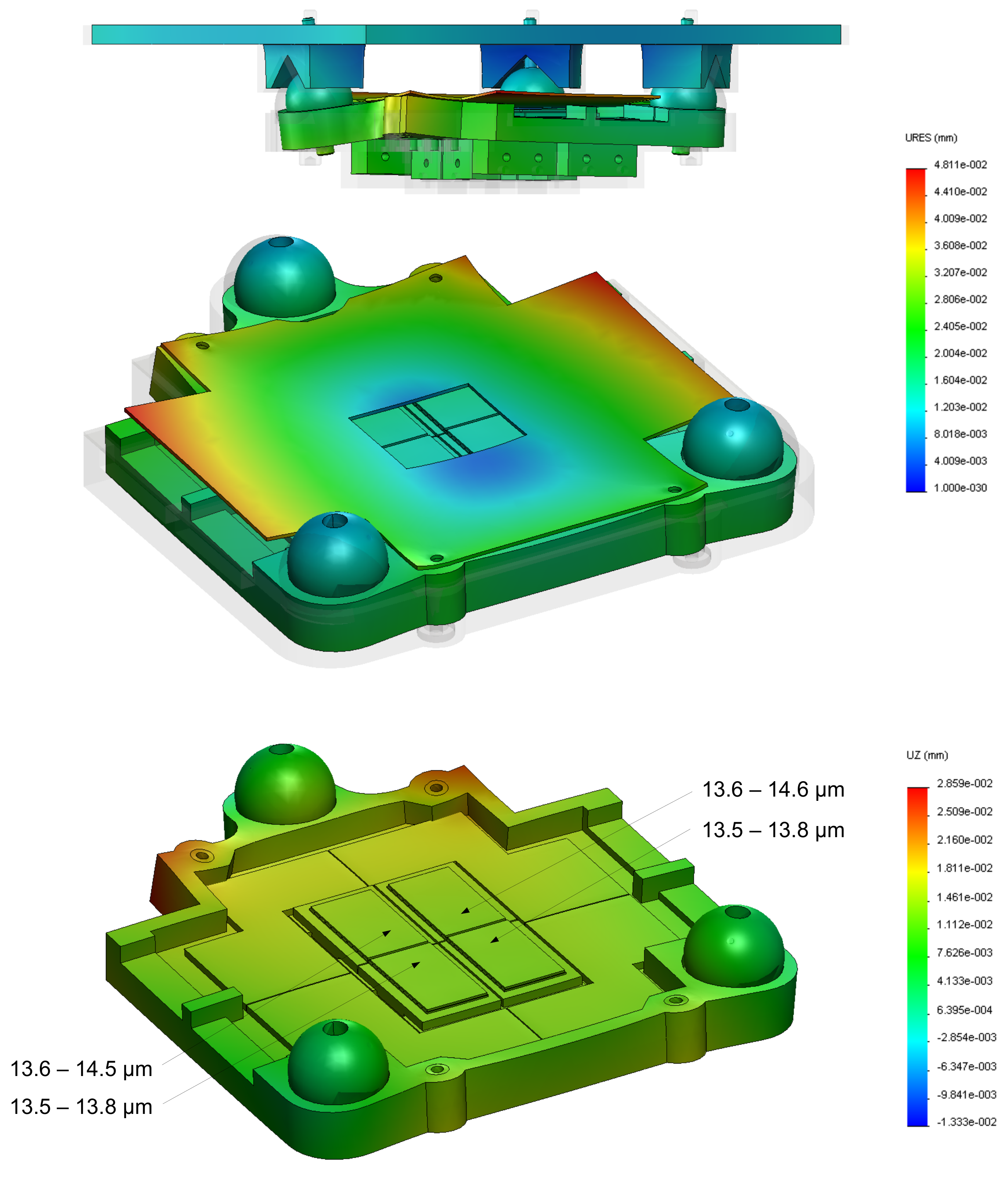}
\caption{\textit{Top:} Worst case estimate of resulting displacement and distortion of the FPA with respect to the telescope dummy plate, illustrated using a scale factor of 250. The translucent image indicates the FPA's position and size at room temperature. The color shades indicate the absolute value of the displacement vector. \textit{Bottom:} Worst case displacement of the CCD's active image zones in direction of the optical axis (\mbox{z-direction}), illustrated using a scale factor of 250. The tungsten cover has been masked out for clarity. The color shades indicate the displacement in \mbox{z-direction}.}
\label{fig:Deformation}
\end{figure}

Figure~\ref{fig:Deformation} provides an estimate of the worst case effect (spheres do not slide in the grooves) of thermally induced stress and strain on the FPA geometry indicating the displacement from the original alignment position at room temperature. Both graphs confirm the idea and feasibility of an athermal design based on the presented material selection. Resulting stresses within the ceramic structure have been assessed as uncritical. Still, the displacement of the active CCD areas can be up to 13.5\,-\,14.6\,\textmu m. This equals already about 50\,\% of the remaining misalignment tolerance of the active image areas of 30\,\textmu m (see Sec.~\ref{sec:requirements}). The remaining tolerance for manufacturing, alignment and telescope-related defocus would therefore only be allowed to be less than about \textpm\,15\,\textmu m in total.

An intentional offset of the active image zones of the detectors from the telescope focal plane during integration at room temperature could compensate the FPA deformation and telescope focal plane shift over temperature and make sure the detectors remain within the allowed alignment tolerance at the target operational temperature in orbit. 

\subsection{Discussion}
The obtained temperature distributions on the FPA are very uniform as it has been expected due to the selection of materials. The size of the radiator panel which is necessary to enable a detector temperature of \mbox{\textless\,-80\,\textdegree C} in the operational hot case dictates the temperature levels on the FPA in all load cases. The thermal simulation illustrates the harsh temperature regime the FPA will face in case no power is available to operate the heaters which are attached to the FPA. In this non-operational cold case, the initially set non-operating temperature limit of \mbox{-95\,\textdegree C} (Sec.~\ref{sec:requirements}) cannot be met. The resulting worst-case thermal environment for the PEM is severe. This can reduce its long-term reliability or even impose the risk of irreversible damage. This remaining risk will be mitigated by implementing redundant heaters and providing a fail-safe power supply to them. In addition, low-temperature testing and qualification of the thick-film hybrids will be conducted to make sure that packaging and interconnection of the electronic components remain intact. Spacecraft operations need to ensure that enough power is available throughout the whole mission to run these heaters if necessary. This fact gains special importance as the uncertainty of a thermal analysis in \mbox{phase B} is likely on the order of $\pm$\,10\,K due to uncertain boundary conditions.

Further studies will investigate if an active thermal control can provide a desired stabilization of the EMCCD temperature of about $\pm$\,2\,K.

In terms of contact conductance, probably the most critical value which needs further investigation is the thermal resistance of the telescope interface. This heat path is critical for thermal control and sizing of the radiator, and estimates of the conductance of the ceramic-metal contact and bolt need to be confirmed or improved. It is therefore recommended to build a representative ground model of the telescope flange and FPA and measure the temperature gradient among the Maxwell mount between the telescope flange demonstrator and base plate to improve the accuracy of future thermal models.

\section{Future work}
\label{sec:approach}

For the concept study presented in this work, it has been assumed that each detector is mounted on an individual ceramic substrate. This implies that all four sensors are separately adjusted (e.g. using precisely ground shims) using a common three point mount which consists of three threaded Invar or Kovar studs (size M3) brazed directly on the ceramic package. It has recently been decided that the FPA design will be continued with a monolithic detector carrier which carries all four EMCCD in order to simplify detector alignment efforts, mitigate array planarity issues and reduce cost.

Current action items for the AsteroidFinder instrument team are the finalization of the CCD package design and adaptation of the presented concept to the final CCD package parameters. The thermal analysis needs to be iterated as soon as boundary conditions are better constrained. Given the usual uncertainties involved in thermal analysis such as unknown contact conductance and environmental heat loads from neighboring subunits, this work can only present a preliminary estimate of the thermal environment on the FPA. For this proof-of-concept study, it only appeared meaningful to study the steady-state case. A remaining action item is a transient thermal analysis which would provide further insights into the thermal environment. 

A mechanical shutter which is necessary to safeguard the CCDs from accidentally looking into the Sun is also in development, but was not covered in this paper.

Work on the AsteroidFinder/SSB FPA is in progress, and several parameters are currently undergoing additional iterations towards a freeze of the presented concept at the end of \mbox{phase B} and a more detailed analysis and design in \mbox{phase C}. The obtained preliminary results provide confidence that the developed concept will be successful. Despite of the somehow limited control on the FE model, the integrated modeling approach has been proven as very helpful for the assessment of the feasibility of an early design and a rapid development.

\section{Conclusions}
\label{sec:conclusions}

A machinable AlN-BN ceramic composite has been selected for the FPA's main structure due to its excellent match to the coefficient of thermal expansion of the telescope and CCD subcarrier material and its high thermal conductivity preventing temperature gradients. This avoids thermally induced stress and distortion, while the manufacturing of the FPA structure is significantly simplified. Consequently, a design in accordance to the special characteristics of ceramics, the selected kinematic mount and the initially discussed requirements has been developed, including the definition of the FPA's mechanical interface to the telescope, thermal interface to a spacecraft radiator panel and electrical interface to the front-end electronic. 

To validate the feasibility of the concept, a finite element model has been implemented in SolidWorks Simulation 2009 and used both for a first steady-state thermal analysis and a related investigation of thermally-induced stress and strain. It became clear that regardless of the athermal design, the thermally-induced displacement of the FPA is a critical item which can affect optical performance. Still, the dimensional stability of the FPA over the very large temperature range from room temperature to operation is considered to be very good. The applied integrated modeling approach using finite elements has proven to be a valuable tool for rapid design iteration and assessment.


\section*{Acknowledgements}
The author greatly acknowledges funding of this work by the German Aerospace Center (DLR), Institute of Planetary Research, Berlin as well as countless fruitful discussions with the members of the AsteroidFinder instrument team and other colleagues at DLR Berlin. Various helpful inputs and data have been provided by Michael Hartl (Kayser-Threde GmbH, Munich), Mathias Kr\"odel (ecm, Munich), Volodymyr Baturkin (DLR Berlin), Roger Morell (NPL, UK), Roman Schmidt, Arne Wollmann (Elbau GmbH, Berlin) and Christian Haberstroh (TU-Dresden). 


\end{document}